\documentclass[longauth]{aa}

\usepackage{graphicx}
\usepackage{longtable}
\usepackage{comment}
\usepackage{txfonts}
\usepackage{color}

\let\ACMmaketitle=\maketitle
\renewcommand{\maketitle}{\begingroup\let\footnote=\thanks \ACMmaketitle\endgroup}

\begin{document} 

   \title{280 one-opposition near-Earth asteroids \\ 
          recovered by the EURONEAR with the Isaac Newton Telescope
         }
   \titlerunning{280 one-opposition NEAs recovered by the EURONEAR with the INT}
   \author{
             O.~Vaduvescu~\inst{1,2,3}\fnmsep\thanks{email: ovidiu.vaduvescu@gmail.com}
          \and
             L.~Hudin~\inst{4}
          \and
             T.~Mocnik~\inst{1}
          \and
             F.~Char~\inst{5}
          \and
             A.~Sonka~\inst{6}
          \and
             V.~Tudor~\inst{1}
          \and
             I.~Ordonez-Etxeberria~\inst{1,7}
          \and
             M.~D\'iaz~Alfaro~\inst{1,8}
          \and
             R.~Ashley~\inst{1}
          \and
             R.~Errmann~\inst{1}
          \and
             P.~Short~\inst{1}
          \and
             A.~Moloceniuc~\inst{9}
          \and
             R.~Cornea~\inst{9}
          \and
             V.~Inceu~\inst{10}
          \and
             D.~Zavoianu~\inst{11}
          \and
             M.~Popescu~\inst{6,12}
          \and
             L.~Curelaru~\inst{9}
          \and
             S.~Mihalea~\inst{9}
          \and
             A.-M.~Stoian~\inst{13}
          \and
             A.~Boldea~\inst{14,15}
          \and
             R.~Toma~\inst{16,9}
          \and
             L.~Fields~\inst{16}
          \and
             V.~Grigore~\inst{9}
          \and
             H.~Stoev~\inst{1}
          \and
             F.~Lopez-Martinez~\inst{1,17}
          \and
             N.~Humphries~\inst{1}
          \and
             P.~Sowicka~\inst{1,18}
          \and
             Y.~Ramanjooloo~\inst{1}
          \and
             A.~Manilla-Robles~\inst{1}
          \and
             F.~C.~Riddick~\inst{1}
          \and
             F.~Jimenez-Lujan~\inst{1}
          \and
             J.~Mendez~\inst{1}
          \and
             F.~Aceituno~\inst{19}
          \and
             A.~Sota~\inst{19}
             D.~Jones~\inst{2,3}
          \and
             S.~Hidalgo~\inst{2,3}
          \and
             S.~Murabito~\inst{2,3}
          \and
             I.~Oteo~\inst{20,21}
          \and
             A.~Bongiovanni~\inst{2,3}
          \and
             O.~Zamora~\inst{2,3}
          \and
             S.~Pyrzas~\inst{2,3,22}
          \and
             R.~G\'enova-Santos~\inst{2,3}
          \and
             J.~Font~\inst{2,3}
          \and
             A.~Bereciartua~\inst{2,3}
          \and
             I.~Perez-Fournon~\inst{2,3}
          \and
             C.~E.~Mart{\'i}nez-V{\'a}zquez~\inst{2,3}
          \and
             M.~Monelli~\inst{2,3}
          \and
             L.~Cicuendez~\inst{2,3}
          \and
             L.~Monteagudo~\inst{2,3}
          \and
             I.~Agulli~\inst{2,3}
          \and
             H.~Bouy~\inst{23,24}
          \and
             N.~Hu\'elamo~\inst{24}
          \and
             M.~Mongui{\'o}~\inst{25}
          \and
             B.~T.~G\"ansicke~\inst{26}
          \and
             D.~Steeghs~\inst{26}
          \and
             N.~P.~Gentile-Fusillo~\inst{26}
          \and
             M.~A.~Hollands~\inst{26}
          \and
             O.~Toloza~\inst{26}
          \and
             C.~J.~Manser~\inst{26}
          \and
             V.~Dhillon~\inst{27,2}
          \and
             D.~Sahman~\inst{27}
          \and
             A.~Fitzsimmons~\inst{28}
          \and
             A.~McNeill~\inst{28}
          \and
             A.~Thompson~\inst{28}
          \and
             M.~Tabor~\inst{29}
          \and
             D.~N.~A.~Murphy~\inst{30}
          \and
             J.~Davies~\inst{31}
          \and
             C.~Snodgrass~\inst{32}
          \and
             A.~H.M.J.~Triaud~\inst{33}
          \and
             P.~J.~Groot~\inst{34}
          \and
             S.~Macfarlane~\inst{34}
          \and
             R.~Peletier~\inst{35}
          \and
             S.~Sen~\inst{35}
          \and
             T.~\.{I}kiz~\inst{35}
          \and
             H.~Hoekstra~\inst{36}
          \and
             R.~Herbonnet~\inst{36}
          \and
             F.~K\"{o}hlinger~\inst{36}
          \and
             R.~Greimel~\inst{37}
          \and
             A.~Afonso~\inst{38}
          \and
             Q.~A.~Parker~\inst{39,40}
          \and
             A.K.H.~Kong~\inst{41}
          \and
             C.~Bassa~\inst{42}
          \and
             Z.~Pleunis~\inst{43}
          }

   \institute{\tiny{\textsuperscript{1} \hspace{0.5 mm} Isaac Newton Group of Telescopes (ING), Apto. 321, E-38700 Santa Cruz de la Palma, Canary Islands, Spain
         \and
             Instituto de Astrof\'{i}sica de Canarias (IAC), C/V\'{i}a L\'{a}ctea s/n, 38205 La Laguna, Tenerife, Spain
         \and
             Departamento de Astrof\'{i}sica, Universidad de La Laguna, 38206 La Laguna, Tenerife, Spain
         \and
             Amateur astronomer, ROASTERR-1 Observatory, 400645 Cluj Napoca, Romania
         \and
             Unidad de Astronom\'ia, Facultad Ciencias B\'asicas, Universidad de Antofagasta, Chile
         \and
             Astronomical Institute of the Romanian Academy, 5 Cutitul de Argint, 040557, Bucharest, Romania
         \and
             Dpto. de F\'isica Aplicada I, Escuela de Ingenier\'ia de Bilbao, Universidad del Pa\'is Vasco, Bilbao, Spain
         \and
             National Solar Observatory, 3665 Discovery Drive, Boulder, CO 80303, USA
         \and
             Romanian Society for Meteors and Astronomy (SARM), Str. Tineretului 1, 130029 Targoviste, Romania
         \and
             Amateur Astronomer, Cluj Napoca, Romania
         \and
             Bucharest Astroclub, B-dul Lascar Catargiu 21, sect 1, Bucharest, 010662, Romania
         \and
             Institut de M\'ecanique C\'eleste et de Calcul des \'Eph\'em\'erides (IMCCE) CNRS-UMR8028, Observatoire de Paris, 77 avenue Denfert-Rochereau, 75014 Paris Cedex, France
         \and
             Amateur astronomer, Schela Observatory, 800259 Schela, Romania
         \and
             Faculty of Sciences, University of Craiova, Str. Alexandru Ioan Cuza 13, 200585 Craiova, Romania
         \and
             Horia Hulubei National Institute for R\&D in Physics and Nuclear Engineering (IFIN-HH), Str. Reactorului 30, Magurele, Romania
         \and
             Armagh Observatory and Planetarium, College Hill, Armagh, BT61 9DG, Northern Ireland
         \and
             Instituto de Astrofísica e Ci\^encias do Espa\c{c}o, Universidade do Porto, CAUP, Rua das Estrelas, 4150-762, Porto, Portugal
         \and
             Nicolaus Copernicus Astronomical Center, Bartycka 18, PL-00-716 Warsaw, Poland
         \and
             Instituto de Astrof\'isica de Andaluc\'ia (IAA-CSIC), Glorieta de la Astronom\'ia, S/N, Granada, 18008, Spain
         \and
             Institute for Astronomy, University of Edinburgh, Royal Observatory, Blackford Hill, Edinburgh EH9 3HJ
         \and
             European Southern Observatory, Karl-Schwarzschild-Str. 2, 85748 Garching, Germany
         \and
             Qatar Environment and Energy Research Institute (QEERI), HBKU, Qatar Foundation, P.O. Box 5825, Doha, Qatar
         \and
             Laboratoire d'astrophysique de Bordeaux, Univ. Bordeaux, CNRS, B18N, all\'ee Geoffroy Saint-Hilaire, 33615 Pessac, France
         \and
             Centro de Astrobiolog\'{\i}a (INTA-CSIC), Dpto. de Astrof\'{\i}sica, ESAC Campus, Camino bajo del Castillo s/n, 28692 Villanueva de la Ca\~nada, Madrid, Spain
         \and
             Centre for Astrophysics Research, Science and Technology Research Institute, University of Hertfordshire, College Lane, Hatfield, AL10 9AB, UK
         \and
             Department of Physics, University of Warwick, Coventry CV4 7AL, UK
         \and
             Department of Physics and Astronomy, University of Sheffield, Sheffield S3 7RH, UK
         \and
             Astrophysics Research Centre, School of Mathematics and Physics, Queen's University Belfast, BT7 1NN, UK
         \and
             School of Physics and Astronomy, University of Nottingham, University Park, Nottingham NG7 2RD, UK
         \and
             Institute of Astronomy, University of Cambridge, Madingley Road, Cambridge CB3 0HA, UK
         \and
             UK Astronomy Technology Centre, Blackford Hill, Edinburgh, EH9 3HJ, Scotland
         \and
             School of Physical Sciences, The Open University, Milton Keynes, MK7 6AA, UK
         \and
             Institute of Astronomy, University of Cambridge, Cambridge CB3 0HA, UK
         \and
             Department of Astrophysics/IMAPP, Radboud University, P.O. Box 9010, 6500 GL, Nijmegen, The Netherlands 
         \and
             Kapteyn Astronomical Institute, University of Groningen, Postbus 800, NL-9700 AV Groningen, the Netherlands
         \and
             Leiden Observatory, Leiden University, PO Box 9513, NL-2300 RA Leiden, the Netherlands
         \and
             Department for Geophysics, Astrophysics and Meteorology, Institute of Physics, NAWI Graz, Universitätsplatz 5, A-8010 Graz, Austria
         \and
             Instituto de Astrofísica e Ci\^encias do Espa\c{c}o, Faculdade de Ci\^encias da Universidade de Lisboa, Portugal
         \and
             The University of Hong Kong, Department of Physics, Hong Kong SAR, China
         \and
             The Laboratory for Space Research, The University of Hong Kong, SAR, China
         \and
             Institute of Astronomy and Department of Physics, National Tsing Hua University, Hsinchu 30013, Taiwan
         \and
             ASTRON, the Netherlands Institute for Radio Astronomy, Postbus 2, NL-7990 AA Dwingeloo, The Netherlands
         \and
             Department of Physics and McGill Space Institute, McGill University, 3600 University Street, Montreal, QC H3A 2T8, Canada
             }
}

   \date{Submitted to A\&A 28 Aug 2017; Re-submitted 10 Oct 2017; Accepted 11 Oct 2017; DOI 10.1051/0004-6361/201731844}


  \abstract
   {One-opposition near-Earth asteroids (NEAs) are growing in number, and they 
    must be recovered to prevent loss and mismatch risk, and to improve their orbits, 
    as they are likely to be too faint for detection in shallow surveys at future 
    apparitions. } 
   {We aimed to recover more than half of the one-opposition NEAs recommended for 
    observations by the Minor Planet Center (MPC) using the Isaac Newton Telescope 
    (INT) in soft-override mode and some fractions of available D-nights. 
    During about 130 hours in total between 2013 and 2016, we targeted 368 
    NEAs, among which 56 potentially hazardous asteroids (PHAs), observing 437 
    INT Wide Field Camera (WFC) fields and recovering 280 NEAs ($76\%$ of all targets). }
   {Engaging a core team of about ten students and amateurs, we used the THELI, 
    Astrometrica, and the Find\_Orb software to identify all moving 
    objects using the blink and track-and-stack method for the faintest targets 
    and plotting the positional uncertainty ellipse from NEODyS. }
   {Most targets and recovered objects had apparent magnitudes centered 
    around $V\sim22.8$ mag, with some becoming as faint as $V\sim24$ mag. 
    One hundred and three objects (representing $28\%$ of all targets) were 
    recovered by EURONEAR alone by Aug 2017. 
    Orbital arcs were prolonged typically from a few weeks to a few years; 
    our oldest recoveries reach 16 years. 
    The O-C residuals for our 1,854 NEA astrometric positions 
    show that most measurements cluster closely around the origin. 
    In addition to the recovered NEAs, 22,000 positions of about 3,500 known 
    minor planets and another 10,000 observations of about 1,500 unknown objects 
    (mostly main-belt objects) were promptly reported to the MPC by our team. 
    Four new NEAs were discovered serendipitously in the analyzed fields and were 
    promptly secured with the INT and other telescopes, while two more NEAs were 
    lost due to extremely fast motion and lack of rapid follow-up time. 
    They increase the counting to nine NEAs discovered by the EURONEAR 
    in 2014 and 2015.
   }
   {Targeted projects to recover one-opposition NEAs are efficient in
    override access, especially using at least two-meter class and preferably larger 
    field telescopes located in good sites, which appear even more efficient than 
    the existing surveys. }

   \keywords{}

   \maketitle
%

\section{Introduction}

The \textup{recovery} of an asteroid is defined as an observation made during a 
second apparition (best-visibility period, which typically takes place around a 
new opposition) following the discovery \citep{boa00}. 
The recovery of poorly observed asteroids and especially near-Earth asteroids 
(NEAs) and near-Earth objects (NEOs) is a very important task to prevent object 
loss and mispairing, and to improve the orbits and dynamical evolution. 

Very few papers have so far described targeted recovery and follow-up 
programs of NEAs. We mention here the pioneering efforts of \cite{tat94}, 
who used three telescopes in Canada (including the DAO 1.85~m) 
to follow up 38 NEAs and recover 2 NEAs during 1992. 
\cite{boa00} presented some statistics based on a sample of multi-opposition 
NEAs, sorting recoveries into four classes that included new observations and data 
mining of existing image archives and concluding that planning telescope 
observations is the best way to recover NEAs. 
\cite{tic00} and \cite{tic02} presented recoveries of 21 NEAs over four and half 
years (1997-2001) using the 0.57~m telescope at Klet' observatory in Slovakia. 
Since 2002, the follow-up (mainly) and recovery efforts at Klet' have been improved 
through the KLENOT program, using a dedicated 1.06~m telescope equipped with a $33^\prime$ 
square camera. Over six and half years (2002-2008), this program counted more than 
1000 NEA follow-up observations, but only 16 NEA recoveries \citep{tic09}, 
suggesting that larger (preferably at least 2~m class) and larger field 
facilities are needed today for recovery. 

During the past few years, recovery of poorly observed NEAs has become 
essential to confirm the orbits of one-opposition objects that have not been 
observed for years since discovery and very short follow-up (typically only 
a few weeks), some in danger of loss or mispairing with newly discovered NEAs. 

Particular attention should be given when telescope time is scarce, requiring a 
larger aperture, field of view, and mandatory quality control of the astrometry 
and orbital fitting. Within the European Near Earth Asteroids Research (EURONEAR) 
\citep{vad08}, follow-up and recovery have been the main astrometric tools used 
for the orbital amelioration of NEAs, potentially hazardous asteroids (PHAs), 
and virtual impactors (VIs) \citep{bir10,vad11,vad13}. 

Since 2000, A. Milani and his Pisa University team have improved the uncertainty 
models needed to search for poorly observed asteroids (one-opposition with short 
arcs, or asteroids that have not been observed for many years), considering nonlinear error propagation 
models to define the sky uncertainty area, which typically spans an elongated 
ellipse \citep{mil99a,mil10}. 
It is essential to use these theories to recover one-opposition NEAs, and this 
could be easily done today using the ephemerides given by the 
NEODyS server\footnote{http://newton.dm.unipi.it/neodys/index.php?pc=0}
or the \textup{OrbFit} Software Package\footnote{http://adams.dm.unipi.it/orbfit}. 

When we count the entire NEA database as of Aug 2017 (about 16,500 objects with orbital 
arcs expressed in days), about $50\%$ represent one-opposition NEAs (more than 8000 
objects), and this percentage is growing because of the accelerated discovery rate of existing and future surveys. 
A pool of about 400 one-opposition NEAs ($5\%$) brighter than $V<24$ mag 
with solar elongation greater than $60^\circ$ are recommended for observations 
at any particular time by the Minor Planet Centre (MPC) at any particular time in their 
Faint\footnote{http://www.minorplanetcenter.net/iau/NEO/FaintRecovery.html} and 
Bright\footnote{http://www.minorplanetcenter.net/iau/NEO/BrightRecovery.html} 
NEA Recovery Opportunities lists. 
Around opposition, many of these targets escape detection by major surveys because 
they are faint, because the visibility windows
are relatively short, because of fast proper motions, and because
of bright 
Moon and Milky Way interference. 

In 2014, we started a pilot recovery program with the aim to observe 100 
one-opposition NEAs using the 2.5~m Isaac Newton Telescope (INT) accessed during
at most 30 triggers (maximum one hour each available night) through the Spanish TAC ToO 
time (Target of Opportunity or override mode). This program produced some promising 
results (about 40 recoveries during only 15 triggers), nevertheless, some visibility 
windows were lost because telescope access was constrained
to only during the allocated 
Spanish one-third fraction, only when the imaging camera was available, and only 
during dark time. 
During the next three semesters, we multiplied the trigger windows by proposing the 
same program to the other two TACs (UK and Dutch), who have agreed to share 
the load and granted 15-20~h each during each of the next three semesters, but 
mostly in ``soft'' mode (only at the discretion of the observer) and also 
accepting some twilight time (20 min mostly before morning) so
that their own research was not strongly affected. 
The first semester in 2016 concluded with the last Spanish allocation, and by mid-2016, we 
reached the goal of recovering more than half of the one-opposition NEAs recommended for observation by the MPC. 

In this paper we report the achievements of this project, discussing the observing 
methods and findings, and comparing the INT with other facilities used for similar projects. 
In Section~2 we present the planning tools and observations. The data reduction software 
and methods are included in Section~3, the results are presented in Section~4, and we
conclude in Section~5.


\section{Planning and observations}

Here we present the tools we used for planning, 
the facilities, and the observing modes. 

\subsection{Recovery planning tool}

In April 2010, the ``One-opposition NEA Recovery Planning'' 
tool\footnote{http://www.euronear.org/tools/planningmpc.php} was written in 
PHP by Marcel Popescu and Ovidiu Vaduvescu to assist in planning the observations of the 
one-opposition NEAs retrieved from the Faint and Bright Recovery Opportunities 
for NEOs MPC lists. 
The input is the observing night (date) and start hour (UT), the number of steps and 
time separator (typically 1h), selection of the bright or faint MPC lists, 
the MPC observatory code, the maximum observable magnitude for the targets, 
the minimum altitude above horizon, the maximum star density in the field (to avoid the Milky Way), 
the maximum proper motion, and the maximum positional uncertainty (one sigma) as retrieved by 
the NEODyS server. 
The output consists of a few tables (one for each time-step), prioritizing targets 
based on a few observability factors to choose from, such as the apparent magnitude, altitude, 
proper motion, sky plane uncertainty, or taking them all into
account at once. 
Other data listed in the output are the stellar density, the angular distance to the Moon, 
and the Moon altitude and illumination. 

\subsection{INT override observations}

The 2.5\,m Isaac Newton Telescope (INT) is owned by the Isaac Newton Group (ING). It is located 
at 2336\,m altitude at the Roque de los Muchachos Observatory (ORM) on La Palma, Canary 
Islands, Spain. The mosaic Wide Field Camera (WFC) is located at the $F/3.3$ INT prime focus, 
consisting of four CCDs with $2048 \times 4098$ $13.5\,\mu$m pixels each, resulting in a scale of 
$0.33\,^{\prime\prime}$/pixel and a total $34^\prime$ square field with a missing small square 
$12^\prime$ in its NW corner. During all runs, we used the Sloan $r$ filter, which suppresses fringing
and improves the target signal-to-noise ratio (S/N) in the twilight. 
The telescope is capable of tracking at differential rates, and we mostly used tracking at half the NEA 
proper motion in order to obtain a similar measurable trailing effect for both the target and reference 
stars. The INT median seeing is $1.2\,^{\prime\prime}$ , and we typically required an ORM seeing 
monitor limit of $1.5\,^{\prime\prime}$ in order for the triggers to become active. 

In Table~1 we include the observing proposals (all three TACs), the number of executed triggers 
(in bold), and the total granted number of triggers (e.g., {\bf 15}/30 means that 15 triggers were 
executed of a maximum allowed 30). Additionally, available fractions during another nine ING discretionary 
nights (``D-nights'') were used to observe a few dozen targets, involving some ING student observers. 
In total, about 130 INT hours were used for this program. All the observers were invited to become 
coauthors of this paper. 

\setlength{\tabcolsep}{10pt} 
\begin{table*}
\caption{Observing proposals and number of triggers activated (in bold) with the Isaac Newton Telescope (INT)} 
\centering
\begin{tabular}{ r | l r |  l r  |  l r  } 
\hline\hline\noalign{\smallskip}
Semester &    \multicolumn{2}{|c|}{SP TAC}   &  \multicolumn{2}{|c|}{UK TAC} &  \multicolumn{2}{|c}{NL TAC} \\ 
\noalign{\smallskip}\hline\hline\noalign{\smallskip}                 
2014A    &  136-INT09/14A (C136)  & {\bf 15}/30 &                  &             &               &             \\  
2014B    &  088-INT10/14B (C88)   &  {\bf 6}/20 &  I/2014B/02 (P2) & {\bf 10}/20 &               &             \\
2015A    &  033-MULT-2/15A (C33)  &  {\bf 9}/20 &  I/2015A/05 (P5) &  {\bf 1}/20 & I15AN003 (N3) &  {\bf 3}/20 \\
2015B    &  001-MULT-2/15B (C1)   & {\bf 14}/15 &  I/2015B/02 (P2) & {\bf 11}/15 & I15BN001 (N1) &  {\bf 4}/15 \\
2016A    &                        &             &  I/2016A/02 (P2) &  {\bf 6}/10 &               &             \\
\noalign{\smallskip}\hline\hline\noalign{\smallskip}    
\end{tabular}
\end{table*}

For each target field, typically 6-8 consecutive images (up to 15 for very faint targets) were 
acquired with exposures of typically 60-90~s each (up to a maximum 180~s in a few cases), so that the trail 
effect would not surpass twice the seeing value. Considering the WFC readout time 
(49~s in the slow and 29~s in the fast mode used mostly in this project), one target sequence could take between 10 
and 20 minutes, which means that we could accommodate between three and six targets during a one-hour typical override. 
For targets with larger uncertainties ($3\sigma \gtrapprox 600^{\prime\prime}$), we observed two 
or three nearby fields that covered more than one degree along the line of variation. 

\subsection{Other telescopes}

In addition to the INT override program, three other telescopes accessible to EURONEAR were used 
to recover a few targets and a few NEA candidate discoveries for a very limited time (about 10~h in total). 
The first was the 4.2~m F/11 William Herschel Telescope (WHT) at ORM equipped with the ACAM 
imaging camera (circular $8^\prime$ field) during two D-nights testing and twilight time, and 
another four nights when the current observer had his targets at very high airmass. 
The second was the ESA 1~m F/4.4 Observing Ground Station (OGS) equipped with a $45^\prime$ 
square field camera at Tenerife Teide Observatory, used during two nights for the recovery of two target 
NEAs and to secure three of our NEA incidental discoveries. 
Additionally, a third telescope was used to follow up a few NEA candidate discoveries, namely the 
Sierra Nevada Observatory 1.5~m (T150) F/12.5 with the CCDT150 camera $8^\prime$ square field. \\

Table~2 lists the observing log, which includes all the 457 observed fields (437 using 
the INT, 12 using the WHT, and 4 using the OGS). We ordered this table based on the 
asteroid designation (first column), then the observing date (start night), listing the 
apparent magnitude $V$ (according to MPC ephemerides), the proper motion $\mu$ and 
the positional uncertainty of the targets (as shown on the observing date by 
MPC at $3\sigma$ level), the number of acquired images (including nearby fields), and the exposure 
time (in seconds). 
In the last three columns we list the current (Jul 2017) status of the
targets (to be discussed next), the MPS publication that includes our recovery, and some comments 
that can include the PHA classification, other used telescopes (WHT or OGS), the track-and-stack 
technique (TS, whenever used), and other possible external stations (MPC observatory code) and the date 
of later recovery (given only for later recoveries when we were
unable to find the targets or for joined 
simultaneous recoveries). 


\section{Data reduction}

We present next the data reduction software and quality control methods used to find and 
measure the targets. 
Three steps were performed during the day following observations: the 
image reduction and field correction (by one person), the visual search and measurement of the 
target and all other moving objects (known or unknown) appearing in each field (distributing 
the work to a team of a few people), and finally the quality control and reporting of all data 
to MPC (by the project leader). 

\subsection{THELI}

Very accurate astrometry (comparable to or lower than the reference star catalog uncertainty, preferably 
below $0.1^{\prime\prime}$) is essential to correctly link and improve the orbits that have been poorly 
observed in the past, like one-opposition NEAs. Any fast system and prime focus larger field camera 
(such as INT-WFC) provides quite distorted raw astrometry that needs correction in order to be used 
for accurate measurements. We used the GUI 
version\footnote{https://www.astro.uni-bonn.de/theli/gui/index.html} of the THELI software 
\citep{erb05,sch13} to reduce the raw WFC images using the night bias and flat field and to 
resolve the field correction to all four CCDs in each WFC-observed field by using a third-degree 
polynomial distortion model. 
In Figure~1 we include one typical field 
distortion map output of THELI (running Scamp), showing pixel scale differences 
of up to $0.006^{\prime\prime}$ between the center and corners of the WFC field, which can produce errors 
of up to $40^{\prime\prime}$when a simple linear astrometric model is applied. 
For most of the data reduction, we used the PPMXL reference star catalog \citep{roe10}, while 
UCAC4, SDSS-DR9, or USNO-B1 were used when the field identification failed because of a  lack of stars 
or small dithering between frames. 

\begin{figure}
\centering
\includegraphics[width=8cm]{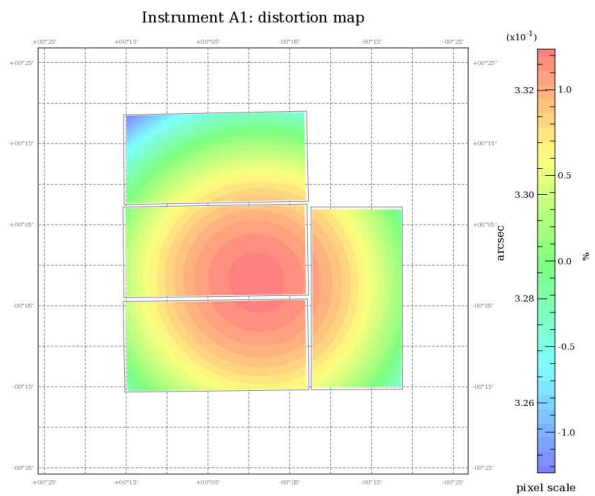}
\caption{Typical THELI field distortion of the INT-WFC field.}
\end{figure}

\subsection{Astrometrica}

The Windows \textup{Astrometrica} software\footnote{http://www.astrometrica.at} is popular among 
amateur astronomers for field registering, object identification, and astrometric measurement of 
the asteroids; it is written by the Austrian amateur astronomer Herbert Raab. We used it after 
every run, updating the MPCORB database to take all newly discovered asteroids
and updated orbits into account. In 2014, Ruxandra Toma and Ovidiu Vaduvescu wrote a user guide 
manual\footnote{http://www.euronear.org/manuals/Astrometrica-UsersGuide-EURONEAR.pdf} (21 pages)
aimed for training the new members of the reduction team. 

\subsubsection{Classic blink search}

We used \textup{Astrometrica} for each observed WFC field to independently blink the four CCDs, 
identifying all moving sources (as known or unknown asteroids), and measuring them. Typically, between 
one and two hours were spent by one reducer for each WFC field. Although \textup{Astrometrica} is capable 
of automatic identification of moving sources, given the faintness of our NEA targets, we decided to use 
visual blink and manual measurements. 
In addition to the targeted NEA, typically up to a few dozen main-belt asteroids (MBAs, about half of 
them known and half unknown) could be identified in good seeing conditions in each observed WFC field.

\subsubsection{Track-and-stack} 

When the NEA target could not be seen using the classic blink search, then the 
\textup{Astrometrica} track-and-stack method (``TS'') was used, either with the ``median'' 
option to eliminate most of the stars, or with the ``add'' option to improve the detection 
of extremely faint targets (S/N=2-3). 
The linear apparent motion assumed by the TS procedure could be affected by the diurnal 
paralax effect, and the TS detection could fail during very close flybys or/and a longer observing 
time that was affected by diurnal effects, but we consider that none of our targets was affected
by these circumstances, as the length of each observing sequence was short. 
To limit the search area, we developed a method using DS9 to load the \textup{Astrometrica} 
TS image and overlay the NEODyS $3\sigma$ uncertainty, thus restraining the visual search 
to a very thin ellipse area (possible to save and load as a DS9 region) typically passing 
across the central CCD4 (holding the target most of the time) or/and nearby CCDs or fields. 
We include in Figure~2 one typical DS9 overlay (NEA 2012~EL5 on 23 Aug 2015 with uncertainty 
$3\sigma=788^{\prime\prime}$ prolonging to the nearby CCD2), which allowed the identification 
of the target falling exactly on the major axis of the NEODyS uncertainty ellipse overlaid
on the stack of $6\times60$s individual images. 

\begin{figure*}
\centering
\includegraphics[width=15cm]{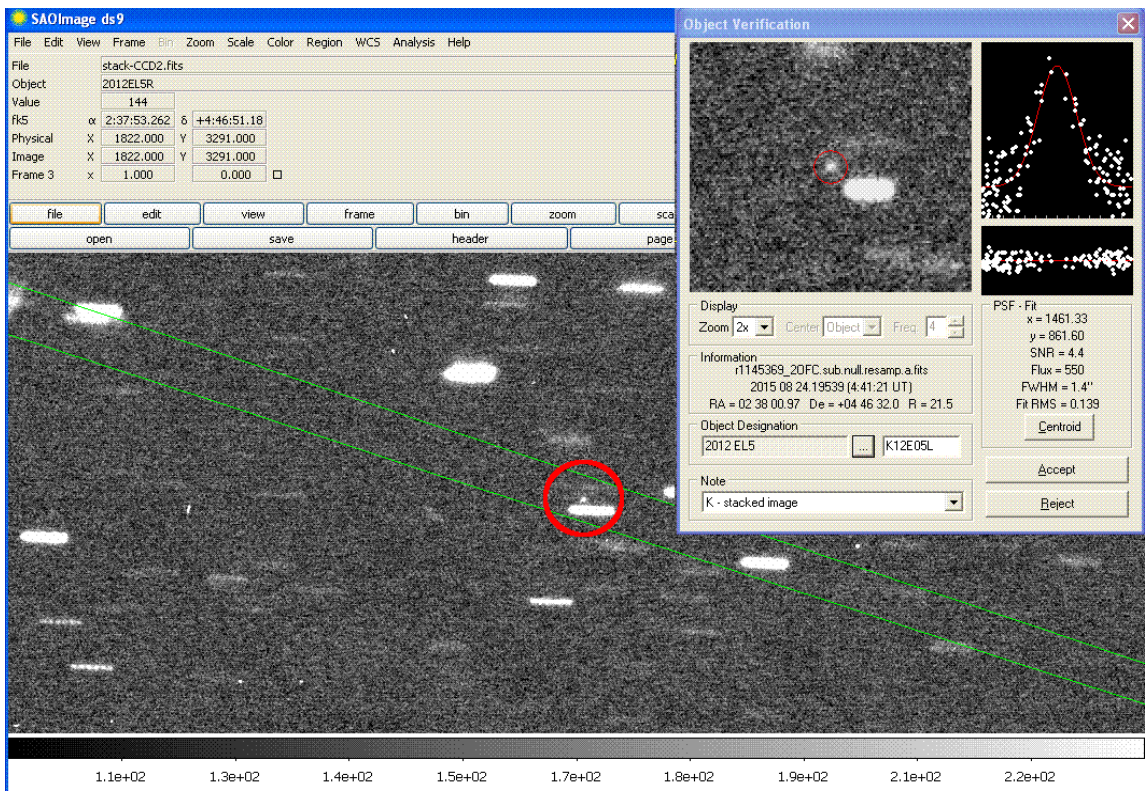}
\caption{Track-and-stack \textup{Astrometrica} image (composition of six individual images
using the "add" option) overlaid on DS9 the NEODyS uncertainty ellipse (green) that we used to 
find the target NEA (2012~EL5, circled in red). }
\end{figure*}

\subsection{Quality control}

\textup{Astrometrica} can easily identify moving sources with well-known asteroids (observed at 
two oppositions at least) by calculating their ephemerides using an osculation orbit model 
with orbital elements very close to the observing time, which provides a very good accuracy of 
$\sim1^{\prime\prime}$. After each observing night, we used \textup{Astrometrica} to check all 
moving sources that were visible in each WFC field against known MBAs included in the updated MPCORB 
database\footnote{http://www.minorplanetcenter.net/iau/MPCORB.html}. 
Nevertheless, one-opposition objects and especially NEAs closer to Earth are affected by 
positional uncertainties, and they should be checked using additional tools. 

\subsubsection{AstroCheck, FITSBLINK, and O-C calculator}

In 2015, Lucian Hudin developed the EURONEAR PHP tool 
\textup{AstroCheck}\footnote{http://www.euronear.org/tools/astchk.php} to verify the consistency 
of all astrometric measurements obtained in each WFC field (known or unknown asteroids). 
This tool assumes that a simple linear regression model holds for relatively short and 
contiguous observational arcs like those observed during 10-20 min runs for each target 
of our recovery project. A maximum error 
(default $0.3^{\prime\prime}$ consistent with WFC pixel size) is allowed, all other outliers 
being flagged in red, these positions being revised or discarded by the reducer. 

We used the server FITSBLINK\footnote{http://www.fitsblink.net/residuals} , which identifies 
known objects and provides tables and plots to check the $O-C$ (observed minus calculated) 
residuals for all asteroids (mostly MBAs) identified by \textup{Astrometrica} in each 
WFC field. 
The calculation of the asteroid positions is based on osculating elements near the current 
running date, so the identification is correct for checks after each observing run, but it 
could fail for older measurements.
The great majority of the residuals are scattered around the origin in the $\alpha-\delta$ 
FITSBLINK plots, proving the correct identification of the MBAs. Some asteroids 
(MBAs and target NEAs) show normal non-systematic clustering around values different 
than zero (typically by a few arcseconds), suggesting the correct identification of poorer 
known orbits. If any object presents systematic O-C residuals (typically located far from 
the origin), then this most probably represents an erroneous identification, and FITSBLINK 
flags these objects as unknown. 

In addition to FITSBLINK, to check MBAs residuals, we used the EURONEAR tool 
\textup{O-C Calculator}\footnote{http://www.euronear.org/tools/omc.php} , which provides tables 
to check for accurate residuals for each target NEA. The residuals are calculated based 
on accurate ephemerides run using the \textup{OrbFit} planetary perturbation model that is 
automatically queried via NEODyS\footnote{http://newton.dm.unipi.it/neodys}. Each 
correctly identified one-opposition NEA target must show normal non-systematic scatter 
(located around a center different than the origin), otherwise the identification is 
false. 

For the target NEAs, the FITSBLINK and the \textup{O-C Calculator} residuals could be 
randomly spread (non-systematic) around a point which may be different than the origin, 
while for most MBAs, the O-C values are typically spread around the origin. 

\subsubsection{Find\_Orb and Orbital Fit}

The \textup{Find\_Orb} software\footnote{https://www.projectpluto.com/find\_orb.htm} is 
a user-friendly popular orbit determination software under Windows or Linux for fitting 
orbits of solar system objects based on existing observations, written by the 
US American amateur astronomer Bill Gray. 
We used \textup{Find\_Orb} to finally check NEA targets that  showed larger positional 
uncertainties. Past observations were downloaded from the MPC Orbits/Observations 
database\footnote{http://www.minorplanetcenter.net/db\_search}, which was updated with our 
proposed identification and astrometric measurements, before using \textup{Find\_Orb} in 
two steps. 

First, using only past positions, 
an orbit is fit in a few (typically 3-4) converging steps by activating all perturbers 
and rejecting outlier measurements greater than $1^{\prime\prime}$ in $\alpha$ or $\delta$. 
Virtually all fits should produce an overall $\sigma$ root-mean-square deviation smaller than 
$1^{\prime\prime}$. 
Second, we append our measurements to the input observation file to load in \textup{Find\_Orb}
to attempt an improved orbital fit in a few (3-4) converging steps, which must conserve or 
slightly improve $\sigma$ (typically by $0.01-0.02^{\prime\prime}$) and show random 
distributions in both $\alpha$ and $\delta$ (typically below $0.3^{\prime\prime}$ in
module) around zero for our measurements. 

If any target presents a systematic $O-C$ trend or increases the $\sigma$ orbital fit, 
then the identification is false or the candidate (typically very faint or found using the
TS technique) represents an artifact and is discarded. 


\section{Results}

\subsection{Targeted NEAs}

We accessed time for the NEA recovery program during 102 nights: 94 nights using
the INT (mostly in override mode for a maximum of 1~h each night and using some D-nights), 
plus another 6 nights using the WHT and 2 nights using the OGS. 
We targeted 368 one-opposition NEAs (including 56 PHAs), observing 453 
fields: 437 with the INT (representing $96\%$ of the program), 12 with 
the WHT, and 4 fields with the OGS. 
We recovered 290 NEAs in total ($79\%$ from all 368 targets), of which 280 targets 
were recovered with the INT. 
One hundred and three recovered objects (representing $28\%$ of all targets) were 
observed at second opposition only by EURONEAR, proving the importance of planned 
recovery compared with shallower surveys. 

Orbital arcs were typically prolonged from a few weeks to a few years, our 
oldest one-opposition recoveries improving orbits of objects that were not seen for up
to 16 years (1999~DB2 and 1999~JO6). Based on Table~2, the user can evaluate 
the extended arc (in years) by simply subtracting the discovery year (first four
digits in the first column NEA designation) from the observing date (first 
four digits standing for the year), the oldest recoveries being included in the 
first part of the table. 

Because they were not recovered during the first attempt, 67 NEAs ($18\%$) were 
targeted multiple (typically two to three) times, some of them  even up to 
six times (2008~ON, resolved during four nights), in order to secure recoveries 
of very faint objects that were seen only with TS and to minimize the risk of false detections. 

We sorted our findings into a few groups that we list in Table~2 under the Status column: 

\begin{itemize}
   \item REC - recovery (followed by other stations); 
   \item RECO - recovery only (not followed by others); 
   \item RECJ - recovery joined (simultaneously with others); 
   \item RECR - revised recovery (in 2017 or following other later recovery); 
   \item NOTF - not found (but found by others later); 
   \item NOTFY - not found yet (by any other station). 
\end{itemize}

We were unable to find 79 objects ($21\%$ of all 368 targets) that are marked 
with status NOTF or NOTFY in Table~2 for several reasons, the most common being that some targets were fainter than 
originally predicted, others were affected by cirrus, calima, or late twilight, and a 
few were hidden by bright stars or have fallen in the WFC gaps. 
Of these, 46 objects ($12\%$) were recovered later by other programs 
or surveys (status NOTF), and another 33 objects ($9\%$) have not been found 
yet (by July 2017); these are marked with the status NOTFY. Additionally, we were able to recover 
16 objects later (status RECR), following a revised search (carried out in 2017) 
based on an orbit that was improved by other programs. 
Here we report the most efficient programs 
(MPC code, facility, and number of later recoveries missed by us): 
568+T12 (CFHT and UH telescopes, 28 recoveries or $7\%$ of all our targets), 
926 (Tenagra II, 9 recoveries), J04 (ESA/OGS Tenerife, 8), F51 (Pan-STARRS 1, 6), 
807+W84 (CTIO and Blanco/DECam, 5), 291 (Spacewatch II, 4), H21 (ARO Westfield, 4), 
705 (SDSS, 3), G96 (Catalina, 2), 695 (KPNO, 2), 033 (KSO, 2), H36 (Sandlot 2), 
675+I41 (Palomar and PTF, 2), 309 (Paranal VLT, 1), G45 (SST Atom Site, 1), and 
T08 (ATLAS-MLO, one recovery). 

In Figure~3 we present the magnitude distribution of all targeted fields (plotted with 
a dotted line) and recovered targets (solid line). Most targets had $V\sim22.8,$ and most 
targets were also recovered around $V\sim22.8$. A few fainter objects were targeted and some were 
recovered close to $V\sim24.0$ using the TS technique. 

\begin{figure}
\centering
\includegraphics[width=8cm]{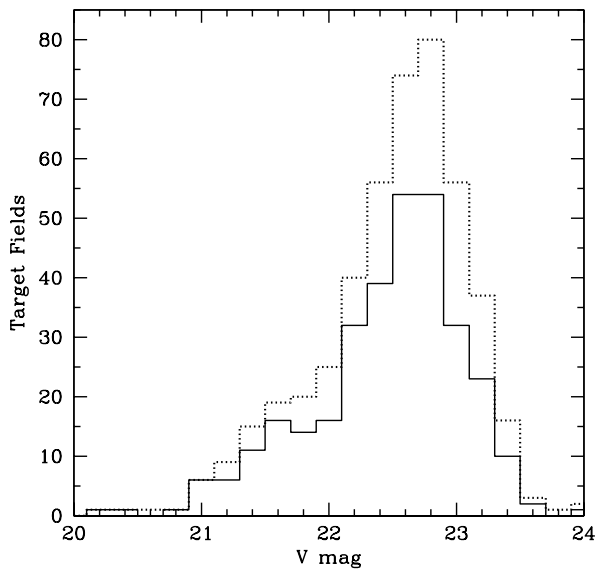}
\caption{Distribution of the NEA apparent magnitude.}
\end{figure}

In Figure~4 we present the proper motion distribution of all targeted fields (dotted line) 
and recovered targets (solid line). Most targets had relatively small proper motion (around 
$\mu\sim0.7^{\prime\prime}$/min, sampling the morning small solar elongation targets), while 
another small peak is visible around $\mu\sim2.0^{\prime\prime}$/min and other faster objects 
(up to $\mu\sim5.0^{\prime\prime}$/min) sample closer flybys and opposition apparitions. 

\begin{figure}
\centering
\includegraphics[width=8cm]{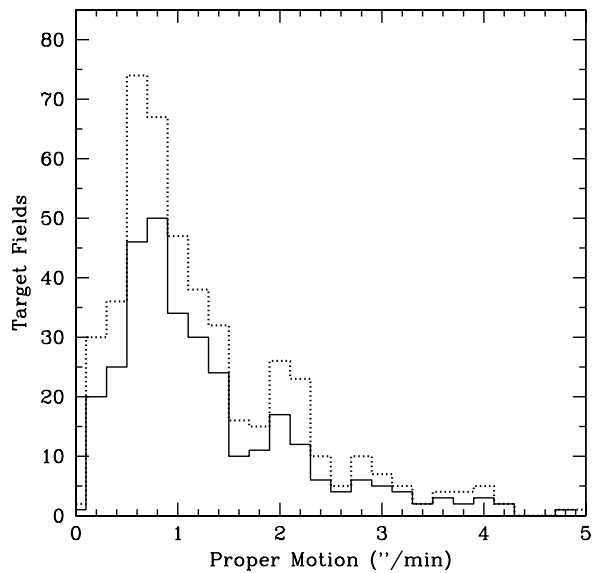}
\caption{Distribution of the NEA proper motion.}
\end{figure}

In Figure~5 we present the $3\sigma$ positional uncertainty distribution of all targeted 
objects (dotted line) and recovered targets (solid line). Most targets had 
$3\sigma < 1000^{\prime\prime}$ (due to the selection limit), and there were 20 targets with 
uncertainties of up to $3000^{\prime\prime}$ (outside the plot) for which we observed two or three nearby fields. 

\begin{figure}
\centering
\includegraphics[width=8cm]{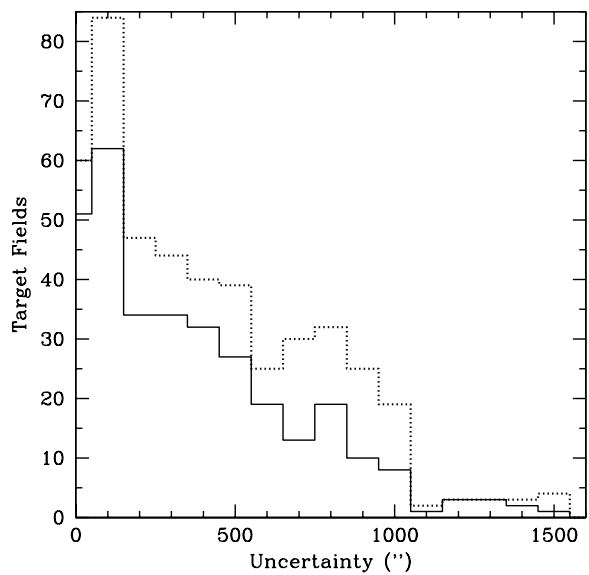}
\caption{Distribution of the NEA $3\sigma$ uncertainty.}
\end{figure}

Figure~6 plots the histogram counting all the observed fields (upper dotted line) 
as a function of the ecliptic latitude ($\beta$), showing that most fields were observed 
between $-20^\circ < \beta < +50^\circ$. The recovered targets are plotted with a 
continuous line (in the middle), and the one-night recoveries are plotted with a 
dashed line (in the bottom). They are discussed in Section~4.2. 

\begin{figure}
\centering
\includegraphics[width=8cm]{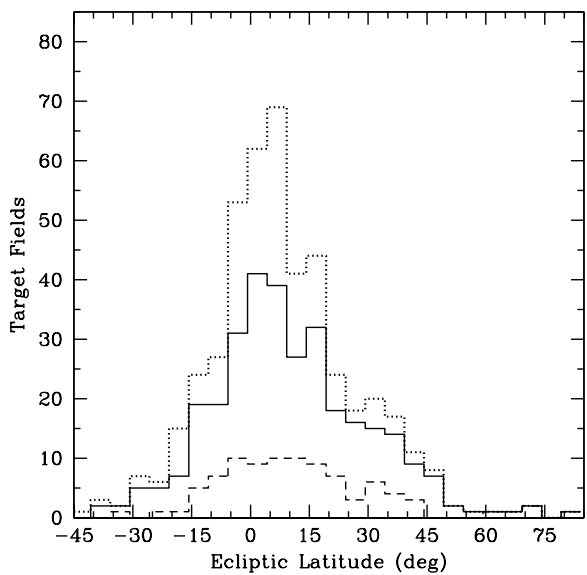}
\caption{Distribution of the NEA ecliptic latitudes.}
\end{figure}

Figure~7 plots the O-C residuals (observed minus calculated) for 1,854 NEA 
measurements from the NEODyS database based on the improved orbits (by 3 Aug 2017). 
Most of the points are located around the origin, with a standard deviation of 
$0.26^{\prime\prime}$ in $\alpha$ and $0.34^{\prime\prime}$ in $\delta$. 
Only eight points ($0.4\%$ of all data) sit outside $1^{\prime\prime}$ 
in either $\alpha$ or $\delta$; they represent measurements of very faint targets. 

\begin{figure}
\centering
\includegraphics[width=8cm]{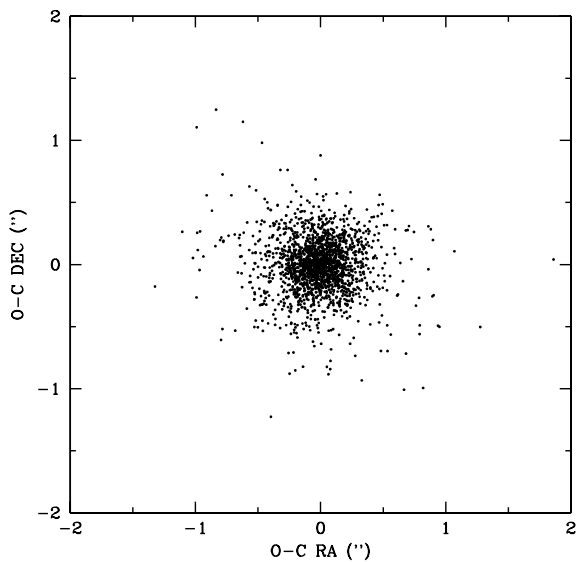}
\caption{O-C residuals for 1,854 positions of 280 one-opposition NEAs.}
\end{figure}

\subsection{Main-belt asteroids and the NEA misidentification risk}

All moving sources found through blinking in the WFC images were identified with known asteroids or were
labeled as unknown asteroids and reported to MPC promptly after each run (typically during 
the next day). By checking the MPCAT-OBS and the ITF 
archives\footnote{http://www.minorplanetcenter.net/iau/ECS/MPCAT-OBS/MPCAT-OBS.html}, 
we were able to count about 22,000 observations of about 3,500 known minor planets (mostly MBAs) 
and about 10,000 observations of about 1,500 unknown objects (most consistent with MBAs) 
reported by our team between Sep 2013 and Oct 2016 as part of this project. 

In a series of papers, A. Milani and colleagues proposed new algorithms to better 
approximate and predict the recovery region of poorly observed asteroids and comets by 
using a nonlinear theory to compute confidence boundaries on the modified target plane 
\citep{mil99a,mil99b,mil00,mil01}. This theory was implemented in NEODyS, which has been 
used by us to plot the uncertainty regions of the one-opposition NEA targets,
which is essential for a correct identification of very faint asteroids (found with the TS 
techique) and one-night recoveries. 
We have made 91 one-night recoveries (counted by Aug 2017), meaning that targets were 
identified and measured during only one night as part of our NEA recovery project (neither 
by us during another night, nor by others until Aug 2017). 
There is some risk for misidentification in these cases when some targets fall in a 
dense ecliptic field populated with MBAs. To assess this risk, in Figure~6 we show 
the ecliptic latitude distribution by plotting all target fields (upper dotted line), 
the recovered target fields (middle solid line), and one-night recoveries (bottom dashed 
line). When we counted the recoveries close to the ecliptic ($-5^\circ < \beta < +5^\circ$), we 
found 19 risk cases ($20\%$ of all one-night recoveries) when target NEAs might be 
confused with MBAs moving at similar direction and rate. The following five precaution 
measures (adopted for most observed fields) minimize false detections in 
these cases: 

\begin{itemize}
   \item We detected all known moving objects and identified all know MBAs and 
         other possible known NEAs in all fields. 
    \item We ensured that O-Cs for the NEA candidate detections were non-systematic 
         (they might spread around a point different than zero, but should not 
         show any systematic trend). 
   \item We plotted the predicted NEODyS uncertainty regions for the target 
         NEAs, ensuring that each candidate NEA detection falls very close to 
         (typically within $1^{\prime\prime}$) the long axis of the NEODyS 
         uncertainty ellipse. 
   \item We fit each candidate detection (positions) to the existing orbit, 
         downloading old observations from the MPC database and using \textup{Find\_Orb} 
         to fit the improved orbit, ensuring that the orbital RMS remains the 
         same or improves slightly (typically by $0.01-0.02^{\prime\prime}$) after 
         the fit and that our candidate positions O-Cs are non-systematic and 
         spread around zero in the new orbital fit. 
   \item We ensured that our measured magnitudes of all targets were similar to 
         their predicted magnitudes (typically within 1 mag, allowing for the  
         unknown color index $r-V$, for some errors in the magnitudes, and for a 
         higher amplitude light-curve that might be due to more elongated objects). 
\end{itemize}

\noindent
Using all these checks, we reduced the risk to confuse any 
one-night target NEA with other MBAs. This is supported 
by many other one-night recoveries that were confirmed later by other 
stations (marked with REC or RECJ in Table~2). 

\subsection{New serendipitous INT NEA discoveries}

\cite{vad15} reported the first EURONEAR NEA discoveries from La Palma that were serendipitously 
found as unknown fast-moving objects in some INT WFC fields taken in 2014 as part of the 
present one-opposition NEA recovery project. 
Here we present discovery circumstances of four other secured NEAs in 2015, plus two 
other probably lost NEOs, together with their composite images shown in Figure~8. 
In total, EURONEAR discovered and secured nine NEAs in 2014 and 2015, the only such findings 
from La Palma and using the INT. 

\begin{figure*}
\centering
\includegraphics[width=18cm]{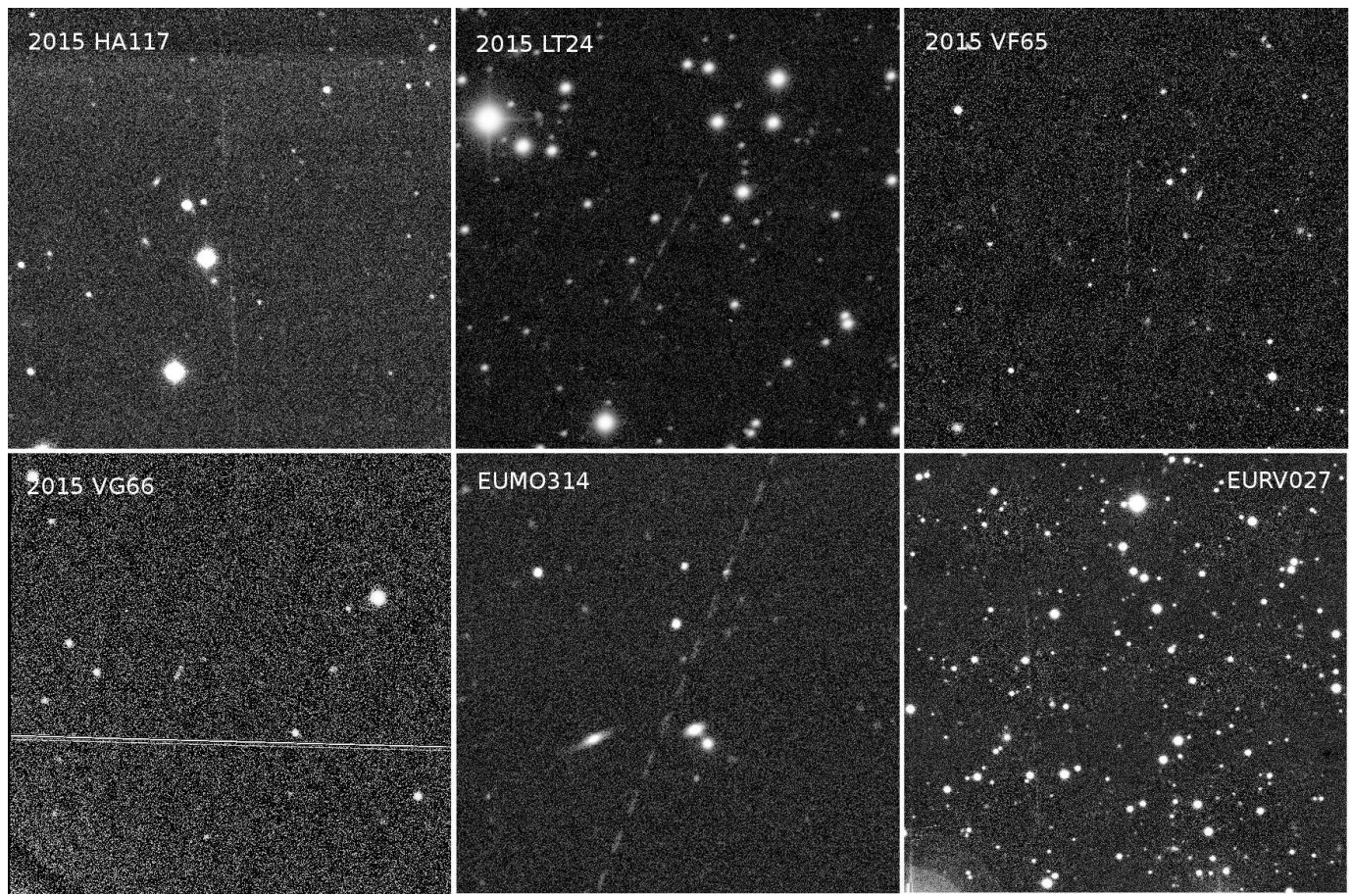}
\caption{Composite images of the six serendipitous NEA discoveries (four secured and two lost objects) 
by EURONEAR using the INT in 2015. Crops are in normal sky orientation (north is up, east to the left), 
$3^\prime\times3^\prime$ field of view, except for EURV027, which is barely visible as three very long 
vertical trails at the left of the $6^\prime\times6^\prime$ field. } 
\end{figure*}

\subsubsection{2015~HA117}

The very fast $15^{\prime\prime}$/min and relatively faint $R\sim22$~mag NEA candidate 
EUHI640 was discovered by Lucian Hudin on 23/24 Apr 2015 in the one-opposition WFC field 
of NEA 2003~WU153 observed by Matteo Monelli and Lara Monteagudo (MPS 603500). 
Thanks to the INT override access, the object was recovered the next night by the same 
team, who scanned 25 WFC fields spanning the MPC uncertainty area, then by the INT, and 
four days later, it was caught by the VLT close to the South celestial pole 
(MPS 604697). 
It became 2015~HA117, estimated at $H=27.2$ and with a size of 10-24~m, apparently having 
an Amor orbit with $MOID=0.01832$~a.u. (based on a seven-day arc), and has 
remained unobserved since then. 

\subsubsection{2015~LT24}

The fast $8^{\prime\prime}$/min EUVI053 $R\sim21.3$~mag NEA candidate was discovered 
by Victor Inceu in images taken on 14/15 Jun 2015 by Stylianos Pyrzas, who chased the
known one-opposition 2012~HO2 NEA (MPS 611632). It was saved during the next night with 
the INT by the same team, then followed-up by other telescopes related to EURONEAR
(OGS 1~m and Sierra Nevada 1.5~m), which prolonged its arc to 11 days. 
Designated as 2015~LT24, it is a relatively large $H=22.4$ 100-220~m Apollo object with 
$MOID=0.15616$~a.u. 

\subsubsection{2015~VF65}

EUHV001 was another very fast ($11^{\prime\prime}$/min) $R\sim22$~mag NEA candidate 
discovered as a trailing object by Lucian Hudin on 7/8 Nov 2015, searching for the 
one-opposition target 2010~VC72 (right field) observed by Odette Toloza (MPS 645818). 
Thanks to the NEOCP posting, it was saved on next night by Spacewatch and the OGS 1~m 
and became 2015~VF65, which was followed-up with the INT and another station later (13 day 
arc). This resolved into an Apollo orbit with an $MOID=0.05225$~a.u. and $H=26.1,$ 
corresponding to a size of 18-40~m. 

\subsubsection{2015~VG66}

EUHV002 was a moderate NEA candidate ($\mu=1.6^{\prime\prime}$/min) relatively 
bright $R=19.4$~mag, first seen on 8/9 Nov 2015 by our most prolific 
discoverer Lucian Hudin in one of the 15 chasing fields (EUHV001I) that were taken
by Odette Toloza to secure our previous NEA candidate (MPS 645822). Despite its 
relatively modest MPC NEO score ($42\%$), we decided to chase it because of its location 
above the NEA border on the $\epsilon-\mu$ plot \citep{vad11}. On the next night, it was 
secured by the INT observers Odette Toloza and Christopher Manser, then precovered 
in Pan-STARRS images by Peter Veres (private communication), and later observed by 
other stations (18-day arc). It has an Apollo orbit with an $MOID=0.01991$~a.u. 
and $H=23.2,$ corresponding to a quite large object of 72-161~m. 

\subsubsection{EUMO314}

This very fast NEA candidate ($\mu=15^{\prime\prime}$/min, $R\sim19.3$~mag) was seen 
by Teo Mocnik in 15 images taken on 1/2 Mar 2015 by Fatima Lopez while chasing another 
faint NEA candidate (EUMO311). It was lost, unfortunately, the WFC being replaced on next 
morning by the IDS spectrograph, while no other station could save it. 

\subsubsection{EURV027}

This extremely fast NEO candidate ($\mu=40^{\prime\prime}$/min) was seen by 
Ovidiu Vaduvescu as four very faint (probably $R\sim23$~mag) and long trails in images 
taken on 14/15 Aug 2015 by Joan Font in the 2013~VM4 target field. This should correspond 
either to a very small (a few meter) object close to opposition or more likely a tiny
geocentric object (Gareth Williams, private communication). It is barely visible 
in Figure~8 as three vertical very long trails on the left side of the composite image. 

\subsubsection{Other NEA candidates}

About 15 other slower ($\mu<1.5^{\prime\prime}$/min) and sometimes extremely faint 
($S/N<5$) NEA candidates were found in some other fields scanned by our program. Most 
of them were chased with the INT on the next nights, and we posted some on the NEOCP list. 
Many of them could not be recovered (even going deeper with the INT), suggesting 
that they are artifacts, while others were recovered and were
found to be MBAs or close NEA species. 
We note the following: 
EUHV056 - a probable Jupiter Trojan ($65\%$, according to MPC), 
2014~RC13 (EUMO201) - Jupiter Trojan (MPO~311499), 
2015~QT4 (EURV028) - Hungaria (MPO~382801), and 
2014~LP9 (EUHT164) - Mars crosser (MPO~300699). 


\section{Conclusions}

A project for recovering one-opposition NEAs recommended by the MPC was carried 
out during a fraction of 102 nights ($\sim130$ hours total) between 2013 and 2016
using the INT telescope equipped with the WFC camera. 
We accessed this time as part of ten proposals with time awarded by three committees 
mostly in soft-override mode and accepting some twilight time, plus other available 
time during a few D-nights. 
The data were rapidly reduced (typically during the next day) by a core team of about 
ten amateurs and students led by the PI, who checked and promptly reported all data 
to the MPC. We outline the following achievements: 

\renewcommand{\labelitemi}{$\bullet$}
\begin{itemize}
\item We targeted 368 one-opposition NEAs (including 56 PHAs)
      for which we observed 437 WFC fields with the INT. 
\item We recovered 290 NEAs ($79\%$ from all targets), 
      sorted into four groups (REC, RECO, RECJ, and RECR), the 
      majority with the INT (280 targets). 
\item Most targets and recovered objects have magnitudes centered around 
      $V\sim22.8$ mag (typically  recovered through blink), while some are 
      as faint as $V\sim24$ mag (only visible with track-and-stack and 
      search in the uncertainty ellipse). 
\item One hundred and three objects ($28\%$ of all targets) have been recovered 
      only by EURONEAR (but no other survey, until Aug 2017 at least). 
\item Orbital arcs were prolonged typically from a few weeks to a few 
      years, our oldest recoveries improving orbits of objects that have not been seen 
      for up to 16 years. 
\item Sixty-seven NEAs ($18\%$) could not be found during a first attempt, and they 
      were targeted multiple (typically two to three) times. 
\item Forty-six objects ($12\%$ of all targets) were not found, but were 
      recovered later by other programs or surveys (UH+CFHT $7\%$, Tenagra, 
      ESA OGS, and major surveys less than $2\%$ of our targets each). 
\item Most targets were slow ($\mu\sim0.7^{\prime\prime}$/min sampling 
      the morning small solar elongation targets), others concentrated 
      around $\mu\sim2.0^{\prime\prime}$/min, while others are faster 
      (up to $\mu=5.0^{\prime\prime}$/min). 
\item Given the WFC $34^\prime$ field, our selection limit in positional 
      uncertainty was $3\sigma < 1000^{\prime\prime}$, but we allowed 20 
      targets with uncertainties up to $3\sigma = 3000^{\prime\prime}$ for 
      which we observed two or three nearby fields. 
\item The O-C residuals for 1,854 NEA measurements show that
most measurements 
      are located closely around the origin, with a standard deviation 
      $0.26^{\prime\prime}$ in $\alpha$ and $0.34^{\prime\prime}$ in $\delta$. 
\item We identified 22,000 observations of about 3500 known minor planets (mostly MBAs) 
      and about 10,000 observations of about 1500 unknown objects (most 
      consistent with MBAs), which were measured and reported to the MPC 
      by our team. 
\item Four new NEAs were discovered serendipitously in the analyzed fields and were
      then secured with the INT and other telescopes, while two more NEAs were 
      lost due to very fast motion and lack of rapid follow-up time. Nine designated 
      NEAs are discovered by the EURONEAR in 2014 and 2015. 
\item Three hundred fifteen MPS publications, including data for one-opposition 
      NEAs, were recovered during this project. 
\end{itemize}

\begin{acknowledgements}
The PI of this project is indebted to the three TACs (Spanish, British, and Dutch) for granting 
INT time (ten proposals during five years) in soft-override mode, which was essential to complete this 
project and secure most discoveries. 
Special thanks are due to M. Micheli (ESA-SSA), observers P. Ruiz, D. Abreu, and the other 
TOTAS team (D. Koschny, M. Busch, A. Kn\"ofel, E. Schwab) for the ESA OGS 1~m follow-up of 
2015~LT24, 2015~VF65, and the attempt to observe 2015~VG66. 
Acknowledgements are due to R. Duffard and S. Martin Ruiz (IAA Granada) for granting some time 
at the Sierra Nevada Observatory (EURONEAR node) with their 1.5~m telescope to secure 2015~LT24 and 2015~VG66. 
Many thanks to O. Hainaut (ESO) and M. Micheli (ESA) for the VLT astrometry of 2015~HA117 (extremely 
fast, faint, and close to the South Pole in just a few days), which prolonged its orbit to a seven-day arc. 
IO acknowledges support from the European Research Council (ERC) in the form of Advanced Grant, {\sc cosmicism}.
RT acknowledges funding for her La Palma trip to Armagh Observatory, which is core-funded by the Northern Ireland Government. 
The research led by BTG, CJM, and NPGF has received funding from the European Research Council under the European
Union's Seventh Framework Programme (FP/2007-2013) / ERC Grant Agreement n. 320964 ({\sc WDTracer}).
Thanks are due to the anonymous referee, whose suggestions helped us to improve the paper. 
\end{acknowledgements}

\newpage
\setlength{\tabcolsep}{5pt} 
\longtab{
\begin{longtable}{lrrrrrrlrl}
\caption{368 NEAs targeted in 453 fields during this project. We explain the columns in Sect. 2.3. }\\
\hline\hline\noalign{\smallskip}
NEA         &    Obs Date &       V &   $\mu$ & 3$\sigma$ &   Imgs &   TEXP &   Status &      MPS &               Comments \\
\noalign{\smallskip}\hline\hline\noalign{\smallskip}
\endfirsthead
\caption{continued.}\\
\hline\hline\noalign{\smallskip}
NEA         &    Obs Date &       V &   $\mu$ & 3$\sigma$ &   Imgs &   TEXP &   Status &      MPS &               Comments \\
\noalign{\smallskip}\hline\hline\noalign{\smallskip}
\endhead
\noalign{\smallskip}\hline\hline\noalign{\smallskip}
\endfoot
\noalign{\smallskip}
    1999DB2 &    20150614 &    23.2 &     0.4 &       277 &      8 &     90 &     NOTF &          &                        \\                   
        ... &    20150818 &    22.6 &     0.7 &       388 &     12 &     90 &     RECO &   621143 &                        \\                   
    1999JO6 &    20150114 &    22.8 &     0.3 &       132 &      6 &    120 &      REC &   561149 &                        \\                   
  2000GV127 &    20140321 &    23.1 &     0.9 &       358 &      8 &    120 &    NOTFY &          &                        \\                   
   2000WY28 &    20140408 &    23.2 &     0.8 &       520 &     10 &    120 &      REC &   511076 &                     TS \\ 
   2001AV43 &    20130907 &    22.0 &     0.4 &       571 &      5 &    180 &      REC &   476154 &                        \\                   
   2001EC16 &    20150114 &    22.6 &     0.8 &        31 &      4 &    120 &      REC &   561150 &                        \\                   
     2001FZ &    20160329 &    23.2 &     0.8 &       763 &      2 &     90 &    NOTFY &          &                        \\
\end{longtable}
}

\end{document}